\documentclass[12pt]{article}
\usepackage{mathtools}
\usepackage{tabularx} 
\usepackage{amsmath}  
\usepackage{graphicx} 
\usepackage[margin=1in]{geometry} 
\usepackage{cite} 
\usepackage[final]{hyperref} 
\hypersetup{
  colorlinks=true,       
  linkcolor=blue,        
  citecolor=blue,        
  filecolor=magenta,     
  urlcolor=blue
}

\begin{document}

\title{$\rho\rho$ scattering revisited with coupled-channels of pseudoscalar mesons}

\date{\today}

\author{Zheng-Li Wang$^{1,2,}$\footnote{Email address:
  \texttt{wangzhengli@itp.ac.cn} }~   and
  Bing-Song Zou$^{1,2,3}$\footnote{Email address:
  \texttt{zoubs@itp.ac.cn} }
  \\[2mm]
  {\it\small$^1$CAS Key Laboratory of Theoretical Physics, Institute
  of Theoretical Physics,}\\
  {\it\small  Chinese Academy of Sciences, Beijing 100190,China}\\
  {\it\small$^2$University of Chinese Academy of Sciences (UCAS), Beijing 100049, China} \\
  {\it\small$^3$Central South University, Hunan 410083, China} \\
}
\maketitle

\begin{abstract}
  The $\rho\rho$ scattering has been studied by two groups which both claimed a dynamical generated scalar meson, most likely to be $f_0(1370)$.
  Here we investigate the influence of coupled-channels of pseudoscalar mesons, {\sl i.e.}, $\pi\pi$ and $\bar KK$, on this dynamical generated scalar state.
  With the coupled channel effect included, the pole and partial decay widths are found to be more close to PDG values for $f_0(1500)$.
\end{abstract}

\section{Introduction}
The chiral unitary approach, which has made much progress in the study of pseudo-scalar meson-meson~\cite{Oller:1997ti}
and meson-baryon~\cite{Oset:1997it, Oller:2000fj}
interactions, has been used to study the interaction of vector mesons among themselves.
The first such study of the $S$-wave $\rho\rho$ interactions found that the
$f_0(1370)$ and the $f_2(1270)$ could be dynamically generated~\cite{Molina:2008jw}. The
work found that the strength of the attractive interaction in the tensor channel is much stronger than that in
the scalar channel, hence leads to a tighter bound tensor state than the corresponding scalar one.

The work~\cite{Molina:2008jw} based on the assumption that the three momenta of the $\rho$ is negligibly small compared to its
large mass. This assumption was questioned by a recent work~\cite{Gulmez:2016scm} which pointed out the importance of relativistic effect for energies around $f_2(1270)$ well below $\rho\rho$ threshold. The $N/D$ method~\cite{Oller:1998zr,Albaladejo:2011bu,Albaladejo:2012sa,Guo:2013rpa,Oller:2014uxa} was used to get the partial wave amplitudes which result a pole for the scalar state similar to Ref.~\cite{Molina:2008jw} but no pole for any tensor state in contradiction with Ref.~\cite{Molina:2008jw}.  However, this conclusion was not agreed upon by Ref.~\cite{Geng:2016pmf} in which the non-relativistic assumption was dropped by evaluating exactly the loops with full relativistic $\rho$ propagators in solving the B-S equation for $\rho\rho$ scattering. Both scalar state and tensor state associated with $f_0(1370)$ and $f_2(1270)$, respectively, were found in consistence with the conclusion of Ref.~\cite{Molina:2008jw}.

From the studies of above two groups, obviously, for the energies around $f_2(1270)$ well below $\rho\rho$ threshold, there is strong model dependence for the interaction of two far off-mass-shell $\rho$ mesons. For the scalar state closer to the $\rho\rho$ threshold, the two groups got similar result rather model independently. In this paper we shall study the influence of coupled-channels of pseudoscalar mesons, {\sl i.e.}, $\pi\pi$ and $\bar KK$, on this dynamical generated scalar state.
In the $\rho\rho-K\bar{K}$ coupling we consider the case of $K$ and $K^*$ exchange,
while in the $\rho\rho-\pi\pi$ coupling we consider the case of $\pi$ and $\omega$ exchange.

\section{Formalism}
\subsection{$\rho\rho \to \pi\pi$ with $\pi$-exchange}
We investigate the coupled channel effect based on a chiral covariant framework~\cite{Gulmez:2016scm}.
We follow the formalism of the hidden gauge interaction which provides the
$\rho\pi\pi$ coupling by means of the Lagrangian~\cite{Bando:1984ej,Bando:1987br}
\begin{equation}\label{eq:VPP}
  \mathcal{L}_{VPP}=-ig \langle V^\mu [P,\partial_\mu P] \rangle.
\end{equation}
where the symbol $\langle \ldots \rangle$ stands for the trace in the $SU(3)$
space with the coupling constant $g=M_V/2f_\pi$ with $M_V$ the mass of vector meson
and $f_\pi=93MeV$ the pion decay constant. The matrices $V_\mu$ and $P$ are given by
\begin{equation}\label{eq:field}
  V_\mu=
  \begin{pmatrix}
    \frac{\rho_0}{\sqrt{2}}+\frac{\omega}{\sqrt{2}} & \rho^+ & K^{*+} \\
    \rho^- & -\frac{\rho_0}{\sqrt{2}}+\frac{\omega}{\sqrt{2}} & K^{*0} \\
    K^{*-} & \bar{K}^{*0} & \phi
  \end{pmatrix}_\mu,
  P=
  \begin{pmatrix}
    \frac{\pi^0}{\sqrt{2}}+\frac{\eta_8}{\sqrt{6}} & \pi^+ & K^+ \\
    \pi^- & -\frac{\pi^0}{\sqrt{2}}+\frac{\eta_8}{\sqrt{6}} & K^0 \\
    K^- & \bar{K}^0 & -\frac{2\eta_8}{\sqrt{6}}
  \end{pmatrix}.
\end{equation}
To get the three different isospin amplitudes for $\rho\rho \to \pi\pi$ we need the
knowledge of the transitions $\rho^+(p_1)\rho^-(p_2) \to \pi^+(p_3)\pi^-(p_4)$,
$\rho^+(p_1)\rho^-(p_2) \to \pi^0(p_3)\pi^0(p_4)$, etc.

\begin{figure}[ht]
  \centering
  \includegraphics{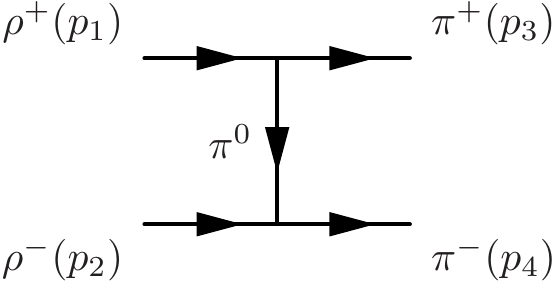}
  \caption{$\pi$-exchange diagram for $\rho^+\rho^- \to \pi^+\pi^-$}
  \label{pi-exchange}
\end{figure}

Starting with the Lagrangian in Eq.\eqref{eq:VPP} we can immediately obtain the amplitude $A_t(p_1,p_2,p_3,p_4)$
of $\rho^+(p_1)\rho^-(p_2) \to \pi^+(p_3)\pi^-(p_4)$ corresponding to Fig.\ref{pi-exchange} as
\begin{equation}
  A_t(p_1,p_2,p_3,p_4)=\frac{-8g^2}{(p_1-p_3)^2-m^2_\pi}
  \epsilon_1 \cdot p_3 \epsilon_2 \cdot p_4.
\end{equation}
In this equation, the $\epsilon_i$ corresponds to the polarization vector of the $i$-th $\rho$.
Each polarization vector is characterized by its three-momentum $\mathbf{p}_i$ and third
component of the spin $\sigma_i$. Explicit expressions of these polarization vectors are
given by \cite{Gulmez:2016scm}
\begin{gather}
  \epsilon(\mathbf{p},0)=
  \begin{pmatrix}
    \gamma \beta \cos \theta \\
    \frac{1}{2}(\gamma-1) \sin 2\theta \cos \phi \\
    \frac{1}{2}(\gamma-1) \sin 2\theta \sin \phi \\
    \frac{1}{2}(1+\gamma+(\gamma-1)\cos 2\theta)
  \end{pmatrix}, \notag \\
  \epsilon(\mathbf{p},\pm1)=\mp \frac{1}{\sqrt{2}}
  \begin{pmatrix}
    \gamma \beta e^{\pm i\phi} \sin \theta \\
    1+(\gamma-1) e^{\pm i\phi} \sin^2 \theta \cos \phi \\
    \pm i+(\gamma-1) e^{\pm i\phi} \sin^2 \theta \sin \phi \\
    \frac{1}{2}(\gamma-1) e^{\pm i\phi} \sin 2\theta
  \end{pmatrix}.
\end{gather}
where $\beta=|\mathbf{p}|/E_p$, $\gamma=1/\sqrt{1-\beta^2}$, $\theta$ and $\phi$ are polar
angle and azimuthal angle of $\mathbf{p}$, respectively. The $u$-channel
$\pi$-exchange amplitude $A_u$ can be obtained from the expression of
$A_t$ by exchanging $p_3 \leftrightarrow p_4$. In this way,
\begin{equation}
  A_u(p_1,p_2,p_3,p_4) = A_t(p_1,p_2,p_4,p_3).
\end{equation}
And now we write the tree-lavel amplitude for $\rho \rho \to \pi \pi$ with $\pi$-exchange
\begin{flalign}\label{eq:amplitudes}
  \rho^+(p_1) \rho^-(p_2) &\to \pi^+(p_3) \pi^-(p_4)  :A_t, & \notag \\
  \rho^+(p_1) \rho^-(p_2) &\to \pi^0(p_3) \pi^0(p_4)  :A_t+A_u, & \notag \\
  \rho^0(p_1) \rho^0(p_2) &\to \pi^+(p_3) \pi^-(p_4)  :A_t+A_u, & \notag \\
  \rho^0(p_1) \rho^0(p_2) &\to \pi^0(p_3) \pi^0(p_4)  :0 &
\end{flalign}

In order to obtain the $S$-wave amplitude in isospin $I=0$ channel we need the isospin
eigenstates. We have
\begin{flalign}\label{eq:isostates}
  \lvert \rho \rho,I=0 \rangle &= -\frac{1}{\sqrt{3}} \lvert \rho^+(p_1) \rho^-(p_2)
  + \rho^-(p_1) \rho^+(p_2) + \rho^0(p_1) \rho^0(p_2) \rangle, & \notag \\
  \lvert \pi \pi,I=0 \rangle &= -\frac{1}{\sqrt{3}} \lvert \pi^+(p_1) \pi^-(p_2)
  + \pi^-(p_1) \pi^+(p_2) + \pi^0(p_1) \pi^0(p_2) \rangle. &
\end{flalign}
where we have used the convention $\lvert \rho^+ \rangle = -\lvert 1,1 \rangle$ and
$\lvert \pi^+ \rangle = -\lvert 1,1 \rangle$ of isospin. By taking into account
Eq.\eqref{eq:isostates} and the amplitudes in Eq.\eqref{eq:amplitudes} we can now
write the isospin $I=0$ amplitude for $\rho\rho \to \pi\pi$
\begin{equation}\label{eq:isoscalar_of_pi}
  T^{(0)}_\pi=16g^2 \left( \frac{\epsilon_1 \cdot p_3 \epsilon_2 \cdot p_4}{m^2_\pi - t}
  + \frac{\epsilon_1 \cdot p_4 \epsilon_2 \cdot p_3}{m^2_\pi - u} \right).
\end{equation}
where $t=(p_1-p_3)^2$ and $u=(p_1-p_4)^2$.

\subsection{$\rho\rho \to \pi\pi$ with $\omega$-exchange}
\begin{figure}[ht]
  \centering
  \includegraphics{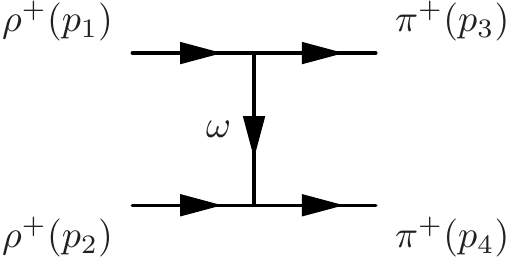}
  \caption{$\omega$-exchange diagram for $\rho^+\rho^- \to \pi^+\pi^-$}
  \label{fig:omega-exchange}
\end{figure}

One needs the $\rho \omega \pi$ coupling which is provided within the
framework \cite{Nagahiro:2008cv} of the hidden gauge formalism by means of the Lagrangian
\begin{equation}\label{eq:VVP}
  \mathcal{L}_{VVP}=\frac{G'}{\sqrt{2}} \epsilon^{\mu \nu \alpha \beta} \langle
  \partial_\mu V_\nu \partial_\alpha V_\beta P \rangle
\end{equation}
with
\begin{equation}
  G'=\frac{3g'^2}{4\pi^2 f_\pi} \qquad g'=-\frac{G_V m_\rho}{\sqrt{2} f_\pi^2}.
\end{equation}
where $G_V=55MeV$ and $f_\pi=93MeV$. At this point we can write down the amplitude of
$\rho^+(p_1) \rho^-(p_2) \to \pi^+(p_3) \pi^-(p_4)$ with $\omega$-exchange corresponding
to Fig.\ref{fig:omega-exchange} as in the $\pi$-exchange case
\begin{equation} \begin{split}
  B_t &= \frac{-G'^2}{(p_1-p_3)^2-m^2_\omega}
  ( p_3 \cdot p_4 p_1 \cdot \epsilon_2 p_2 \cdot \epsilon_1
  + p_1 \cdot p_2 p_4 \cdot \epsilon_1 p_3 \cdot \epsilon_2
  + p_1 \cdot p_4 p_2 \cdot p_3 \epsilon_1 \cdot \epsilon_2 \\
  &- p_2 \cdot p_3 p_1 \cdot \epsilon_2 p_4 \cdot \epsilon_1
  - p_1 \cdot p_4 p_2 \cdot \epsilon_1 p_3 \cdot \epsilon_2
  - p_1 \cdot p_2 p_3 \cdot p_4 \epsilon_1 \cdot \epsilon_2).
\end{split} \end{equation}

And the $u$-channel $\omega$-exchange amplitude $B_u(p_1,p_2,p_3,p_4)$ can be obtained
from the expression of $B_t$ as the case in $\pi$-exchange by exchanging
$p_3 \leftrightarrow p_4$, thus
\begin{equation}
  B_u(p_1,p_2,p_3,p_4)=B_t(p_1,p_2,p_4,p_3).
\end{equation}
Next we write the tree-level amplitude for $\rho\rho \to \pi\pi$ with $\omega$-exchange
\begin{flalign}
  \rho^+(p_1) \rho^-(p_2) &\to \pi^+(p_3) \pi^-(p_4) : B_t, & \notag \\
  \rho^+(p_1) \rho^-(p_2) &\to \pi^0(p_3) \pi^0(p_4) : 0, & \notag \\
  \rho^0(p_1) \rho^0(p_2) &\to \pi^+(p_3) \pi^-(p_4) : 0, & \notag \\
  \rho^0(p_1) \rho^0(p_2) &\to \pi^0(p_3) \pi^0(p_4) : B_t+B_u. &
\end{flalign}
Then using Eq.\eqref{eq:isostates} we can get the $I=0$ amplitude
\begin{equation}\label{eq:isoscalar_of_omega} \begin{split}
  T^{(0)}_\omega &= \frac{-G'^2}{(p_1-p_3)^2-m^2_\omega}
  ( p_3 \cdot p_4 p_1 \cdot \epsilon_2 p_2 \cdot \epsilon_1
  + p_1 \cdot p_2 p_4 \cdot \epsilon_1 p_3 \cdot \epsilon_2
  + p_1 \cdot p_4 p_2 \cdot p_3 \epsilon_1 \cdot \epsilon_2 \\
  &- p_2 \cdot p_3 p_1 \cdot \epsilon_2 p_4 \cdot \epsilon_1
  - p_1 \cdot p_4 p_2 \cdot \epsilon_1 p_3 \cdot \epsilon_2
  - p_1 \cdot p_2 p_3 \cdot p_4 \epsilon_1 \cdot \epsilon_2)
  + (p_3 \leftrightarrow p_4).
\end{split} \end{equation}

\subsection{$\rho\rho \to K\bar{K}$ with $K$-exchange}
\begin{figure}[ht]
  \centering
  \includegraphics{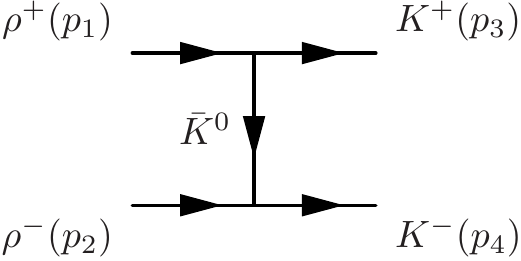}
  \caption{$K$-exchange amplitude for $\rho^+(p_1) \rho^-(p_2) \to K^+(p_3) K^-(p_4)$}
  \label{fig:K-exchange}
\end{figure}

The $\rho KK$ coupling is provided in the same Lagrangian in Eq.\eqref{eq:VPP}, so we can
immediately write down the amplitude of $\rho^+(p_1) \rho^-(p_2) \to K^+(p_3) K^-(p_4)$
with $K$-exchange corresponding to Fig.\ref{fig:K-exchange}
\begin{equation}
  C_t(p_1,p_2,p_3,p_4)= -4g^2 \frac{\epsilon_1 \cdot p_3 \epsilon_2 \cdot p_4}
  {(p_1-p_3)^2-m^2_K}.
\end{equation}
and the $u$-channel
\begin{equation}
  C_u(p_1,p_2,p_3,p_4)=C_t(p_1,p_2,p_4,p_3).
\end{equation}

Then we can obtain the tree-level amplitudes for $\rho\rho \to K\bar{K}$
with $K$-exchange as the following
\begin{flalign}
  \rho^+(p_1) \rho^-(p_2) &\to K^+(p_3) K^-(p_4) : C_t, & \notag \\
  \rho^+(p_1) \rho^-(p_2) &\to K^0(p_3) \bar{K}^0(p_4) : C_u, & \notag \\
  \rho^0(p_1) \rho^0(p_2) &\to K^+(p_3) K^-(p_4) : \frac{C_t+C_u}{2}, & \notag \\
  \rho^0(p_1) \rho^0(p_2) &\to K^0(p_3) \bar{K}^0(p_4) : \frac{C_t+C_u}{2}. &
\end{flalign}

Similar to Eq.\eqref{eq:isostates} we need the isospin $I=0$ eigenstate for
$\lvert K\bar{K} \rangle$. We have
\begin{flalign}\label{eq:isostate}
  \lvert K\bar{K} \rangle = -\frac{1}{\sqrt{2}}
  \lvert K^+(p_1) K^-(p_2) + K^0(p_1) \bar{K}^0(p_2) \rangle. & &
\end{flalign}
where we use the convention $\lvert K^+ \rangle = -\lvert \frac{1}{2},\frac{1}{2} \rangle$
of isospin. By using the isospin wave functions we can obtain for $I=0$
\begin{equation}\label{eq:isoscalar_of_K}
  T^{(0)}_K = 2\sqrt{6} g^2 \left( \frac{\epsilon_1 \cdot p_3 \epsilon_2 \cdot p_4}
  {m^2_K-t} + \frac{\epsilon_1 \cdot p_4 \epsilon_2 \cdot p_3}{m^2_K-u} \right)
\end{equation}
with $t$ and $u$ the usual Mandelstam variable. We can see that the Eq.\eqref{eq:isoscalar_of_K}
is similar to the Eq.\eqref{eq:isoscalar_of_pi}. The former can be obtained from the latter
just by substituting $16 \to 2\sqrt{6}$ and $m_\pi \to m_K$.

\subsection{$\rho\rho \to K\bar{K}$ with $K^*$-exchange}
\begin{figure}[ht]
  \centering
  \includegraphics{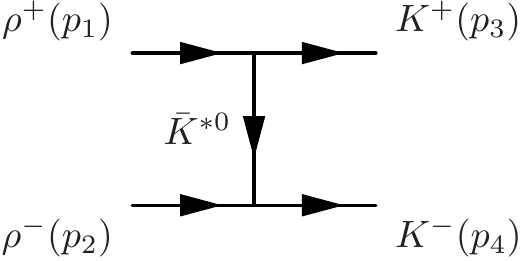}
  \caption{$K^*$-exchange diagram for $\rho^+ \rho^- \to K^+ K^-$}
  \label{fig:Kstar-exchange}
\end{figure}

As for the $\rho K K^*$ coupling, we use the Lagrangian in Eq.\eqref{eq:VVP}. Then we get
the amplitude for $\rho^+(p_1) \rho^-(p_2) \to K^+(p_3) K^-(p_4)$ with $K^*$-exchange
corresponding to the Fig.\ref{fig:Kstar-exchange} as
\begin{equation} \begin{split}
  D_t &= \frac{-G'^2/2}{(p_1-p_3)^2-m^2_{K^*}}
  ( p_3 \cdot p_4 p_1 \cdot \epsilon_2 p_2 \cdot \epsilon_1
  + p_1 \cdot p_2 p_4 \cdot \epsilon_1 p_3 \cdot \epsilon_2
  + p_1 \cdot p_4 p_2 \cdot p_3 \epsilon_1 \cdot \epsilon_2 \\
  &- p_2 \cdot p_3 p_1 \cdot \epsilon_2 p_4 \cdot \epsilon_1
  - p_1 \cdot p_4 p_2 \cdot \epsilon_1 p_3 \cdot \epsilon_2
  - p_1 \cdot p_2 p_3 \cdot p_4 \epsilon_1 \cdot \epsilon_2).
\end{split} \end{equation}
and the $u$-channel
\begin{equation}
  D_u(p_1,p_2,p_3,p_4)=D_t(p_1,p_2,p_4,p_3).
\end{equation}

Next we list the tree-level amplitudes for $\rho\rho \to K\bar{K}$ with $K^*$-exchange as the following:
\begin{flalign}
  \rho^+(p_1) \rho^-(p_2) &\to K^+(p_3) K^-(p_4) : D_t, & \notag \\
  \rho^+(p_1) \rho^-(p_2) &\to K^0(p_3) \bar{K}^0(p_4) : D_u, & \notag \\
  \rho^0(p_1) \rho^0(p_2) &\to K^+(p_3) K^-(p_4) : \frac{D_t+D_u}{2}, & \notag \\
  \rho^0(p_1) \rho^0(p_2) &\to K^0(p_3) \bar{K}^0(p_4) : \frac{D_t+D_u}{2}. &
\end{flalign}
Using Eqs.\eqref{eq:isostates} and \eqref{eq:isostate} we obtain the $I=0$ amplitude
\begin{equation}\label{eq:isoscalar_of_Kstar} \begin{split}
  T^{(0)}_{K^*} &= \frac{-\sqrt{6}G'^2/4}{(p_1-p_3)^2-m^2_{K^*}}
  ( p_3 \cdot p_4 p_1 \cdot \epsilon_2 p_2 \cdot \epsilon_1
  + p_1 \cdot p_2 p_4 \cdot \epsilon_1 p_3 \cdot \epsilon_2
  + p_1 \cdot p_4 p_2 \cdot p_3 \epsilon_1 \cdot \epsilon_2 \\
  &- p_2 \cdot p_3 p_1 \cdot \epsilon_2 p_4 \cdot \epsilon_1
  - p_1 \cdot p_4 p_2 \cdot \epsilon_1 p_3 \cdot \epsilon_2
  - p_1 \cdot p_2 p_3 \cdot p_4 \epsilon_1 \cdot \epsilon_2)
  + (p_3 \leftrightarrow p_4).
\end{split} \end{equation}
which can be obtain from Eq.\eqref{eq:isoscalar_of_omega} by substituting
$1 \to \sqrt{6}/4$ and $m_\omega \to m_{K^*}$.

\subsection{Partial-wave decomposition}
In term of these amplitudes with isospin $I=0$, we can calculating the partial-wave amplitudes
in the $\ell SJI$ basis~ \cite{Gulmez:2016scm} , denoted as $T^{(JI)}_{\ell S;\bar{\ell} \bar{S}}(s)$
for the transition $(\bar{\ell} \bar{S} JI) \to (\ell SJI)$
\begin{equation}\label{eq:partialwaveformula} \begin{split}
  T^{(JI)}_{\ell S;\bar{\ell} \bar{S}}(s) &=
  \frac{Y^0_{\bar{\ell}}(\hat{\mathbf{z}})}{\sqrt{2}^N (2J+1)}
  \sum_{\begin{subarray}{c}
      \sigma_1, \sigma_2, \bar{\sigma}_1 \\ \bar{\sigma}, m
  \end{subarray}}
  \int \mathrm{d} \hat{\mathbf{p}}'' Y^m_\ell(\hat{\mathbf{p}}'')^*
  (\sigma_1 \sigma_2 M \vert s_1 s_2 S) \\
  &\times (m M \bar{M} \vert \ell SJ)
  (\bar{\sigma}_1 \bar{\sigma}_2 \bar{M} \vert \bar{s}_1 \bar{s}_2 \bar{S})
  (0 \bar{M} \bar{M} \vert \bar{\ell} \bar{S} J) T^{(I)}(p_1,p_2,p_3,p_4)
\end{split} \end{equation}
with $M=\sigma_1 + \sigma_2$ and $\bar{M}= \bar{\sigma}_1 + \bar{\sigma}_2$. And $N$
accounts for identical particles, for example
\begin{flalign}
  N&=2 \quad \text{for} \quad \rho\rho \to \pi\pi, & \notag \\
  N&=1 \quad \text{for} \quad \rho\rho \to K\bar{K}. &
\end{flalign}
By using Eq.\eqref{eq:partialwaveformula} we can calculate the partial-wave projected
tree-level amplitudes of Eqs.\eqref{eq:isoscalar_of_pi}, \eqref{eq:isoscalar_of_omega},
\eqref{eq:isoscalar_of_K} and \eqref{eq:isoscalar_of_Kstar} with quantum number
$I,\ell,S=0,0,0$. We denote $T^{(00)}_{00;00}$ by $V$ for simplicity and we have
\begin{flalign}
  &\text{for } \rho \rho \to \pi \pi \text{ with } \pi \text{-exchange} & \notag \\
  &V_\pi = \frac{2g^2}{\sqrt{3}} \left( \frac{2(m^2_\rho-4m^2_\pi)}
  {\sqrt{s-4m^2_\pi} \sqrt{s-4m^2_\rho}} \ln
  \frac{s-2m^2_\rho+\sqrt{s-4m^2_\pi} \sqrt{s-4m^2_\rho}}
  {s-2m^2_\rho-\sqrt{s-4m^2_\pi} \sqrt{s-4m^2_\rho}}
+\frac{s}{m^2_\rho} +2 \right). &
\end{flalign}
\begin{flalign}
  &\text{for } \rho \rho \to \pi \pi \text{ with } \omega \text{-exchange} & \notag \\
  &V_\omega= \frac{G'^2 s}{2\sqrt{3}} \left( \frac{m^2_\omega}
  {\sqrt{s-4m^2_\pi} \sqrt{s-4m^2_\rho}} \ln \frac
  {s+2m^2_\omega-2m^2_\pi-2m^2_\rho+\sqrt{s-4m^2_\pi} \sqrt{s-4m^2_\rho}}
  {s+2m^2_\omega-2m^2_\pi-2m^2_\rho-\sqrt{s-4m^2_\pi} \sqrt{s-4m^2_\rho}}
-1 \right). &
\end{flalign}
\begin{flalign}
  &\text{for } \rho \rho \to K \bar{K} \text{ with } K \text{-exchange} & \notag \\
  &V_K = \frac{g^2}{2} \left( \frac{2(m^2_\rho-4m^2_K)}
  {\sqrt{s-4m^2_K} \sqrt{s-4m^2_\rho}} \ln
  \frac{s-2m^2_\rho+\sqrt{s-4m^2_K} \sqrt{s-4m^2_\rho}}
  {s-2m^2_\rho-\sqrt{s-4m^2_K} \sqrt{s-4m^2_\rho}}
+\frac{s}{m^2_\rho} +2 \right). &
\end{flalign}
\begin{flalign}
  &\text{for } \rho \rho \to K \bar{K} \text{ with } K^* \text{-exchange} & \notag \\
  &V_{K^*}= \frac{G'^2 s}{4} \left( \frac{m^2_{K^*}}
  {\sqrt{s-4m^2_K} \sqrt{s-4m^2_\rho}} \ln \frac
  {s+2m^2_{K^*}-2m^2_K-2m^2_\rho+\sqrt{s-4m^2_K} \sqrt{s-4m^2_\rho}}
  {s+2m^2_{K^*}-2m^2_K-2m^2_\rho-\sqrt{s-4m^2_K} \sqrt{s-4m^2_\rho}}
-1 \right). &
\end{flalign}

\section{Results and discussion}
We label the three channels, $\rho\rho$, $K\bar{K}$ and $\pi\pi$ as 1, 2 and 3, respectively. With the channel transition amplitudes $V_\pi$, $V_\omega$, $V_K$ and $V_{K^*}$ given in last section, we calculate the full amplitude and its pole positions by using the Bethe Salpeter
equation in its on-shell factorized form~\cite{Molina:2008jw,Gulmez:2016scm}
\begin{equation}
  T=\frac{V}{1-VG}.
\end{equation}
$G$ is a diagonal matrix made up by the two-point loop function~\cite{Molina:2008jw,Gulmez:2016scm}
\begin{equation}
  G_{jj}(s)=i\int \frac{d^4q}{(2\pi)^4} \frac{1}{(q^2-m^2_j)((P-q)^2-m^2_j)}
\end{equation}
with $P$ the total four-momentum of the meson-meson systems and $q$ the loop momentum.
The channel is labelled by the subindex $j$. By using dimensional
regularization the integration can be recast as
\begin{equation}
  G_{jj}(s)=\frac{1}{(4\pi)^2} \left( a(\mu) + \log \frac{m^2_j}{\mu^2}
  + \sigma \log \frac{\sigma+1}{\sigma-1} \right)
\end{equation}
with
\begin{equation}
  \sigma=\sqrt{1-\frac{4m^2_j}{s}}
\end{equation}
or using a momentum cutoff $q_\text{max}$ as
\begin{equation}
  G_{jj}(s)=\frac{1}{2\pi^2} \int_0^{q_\text{max}} \mathrm{d} q
  \frac{q^2}{w(s-4w^2+i\epsilon)}
\end{equation}
where $w=\sqrt{q^2+m^2_j}$. The integral can be
done algebraically
\begin{equation}
  G_{jj}(s)=\frac{1}{(4\pi)^2} \left \{ \sigma \log
    \frac{\sigma \sqrt{1+\frac{m^2_j}{q^2_\text{max}}} +1}
    {\sigma \sqrt{1+\frac{m^2_j}{q^2_\text{max}}} -1} -
    2\log \left[ \frac{q_\text{max}}{m_j} \left( 1+ \sqrt{1+ \frac{m^2_j}{q^2_\text{max}}}
  \right) \right] \right \}
\end{equation}

Typical values of the cutoff $q_\text{max}$ are around 1 GeV. $G_{jj}(s)$ has a
right-hand cut above the threshold $2m_j$. In order to make an analytical extrapolation
to second Riemann sheet we make use of the continuity property

\begin{equation}
  G_{jj}^{(2)}(\sqrt{s}+i\epsilon)=G_{jj}(\sqrt{s}-i\epsilon)
\end{equation}
where the index $(2)$ indicates the second Riemann sheet of $G_{jj}$. Then
\begin{align}
  G_{jj}^{(2)}(\sqrt{s}+i\epsilon) &= G_{jj}(\sqrt{s}-i\epsilon)=
  G_{jj}(\sqrt{s}+i\epsilon) - 2i\text{Im}G_{jj}(\sqrt{s}+i\epsilon) \notag \\
  &= G_{jj}(\sqrt{s}+i\epsilon) + \frac{i}{4\pi} \frac{|\mathbf{p}|}{\sqrt{s}}
\end{align}

Other potential of coupled-channels like $\pi\pi-K\bar{K}$ can be found in \cite{Oller:1997ti}.
Our results are shown in Table 1 for various $q_\text{max}$ values.  For comparison, the results for the $\rho\rho$ single channel without considering the coupled channel effects as in Ref.~\cite{Gulmez:2016scm} are show in the second row. The $3\sim 6$ rows show the results including one coupled channel with the exchanged meson listed in the first column. For example the $\pi$ denotes the $\rho\rho-\pi\pi$ channel with $\pi$ exchange and so on. The 7-th row gives the results including all three coupled channels of $\rho\rho$, $\pi\pi$ and $\bar KK$.

\begin{table}[ht]
  \centering
  \[ \begin{array}{|c|c|c|c|c|}
      \hline
      q_{max}(GeV) & 0.875 & 1.0 & 1.2 & 1.4 \\
      \hline
      \rho\rho \text{ only} & 1494.8 & 1467.2 & 1427.3 & 1395.0\\
      \hline
      \pi & 1530.0-4.9i & 1519.5-8.4i & 1501.5-12.3i & 1488.6-14.6i\\
      \hline
      \omega & 1492.2-0.7i & 1466.5-1.0i & 1428.1-1.1i & 1400.0-1.1i\\
      \hline
      K & 1497.8-3.3i & 1473.9-4.1i & 1437.2-4.4i & 1410.0-4.2i\\
      \hline
      K^* & 1489.6-0.5i & 1463.3-0.5i & 1424.5-0.4i & 1396.1-0.3i\\
      \hline
      \text{3-channels} & 1529.8-4.9i & 1519.0-8.6i & 1500.9-13.5i & 1488.4-16.7i\\
      \hline
  \end{array} \]
  \caption{Pole position for coupled-channels}
\end{table}

The above results show that the influence of vector meson $\omega$ and $K^*$ exchanges is very small; the largest influence comes from the $\rho\rho-\pi\pi$ channel coupling by the pion exchange, which shifts up the pole mass and results in a sizable $\pi\pi$ decay width, comparable with relevant PDG values for $f_0(1500)$~\cite{Tanabashi:2018oca}.  For the $\rho\rho-K\bar{K}$ coupled-channel case we can see that the width is
consistent with $f_0(1500)$ decaying into $K\bar{K}$ in PDG, which is about $8.9MeV$.
When taking into account all the three channels, the pole position is close to the results by considering only the pion exchange contribution.
With $q_\text{max}=1.4GeV$, the pole mass and partial decay widths to $\pi\pi$ and $\bar KK$ are roughly consistence with PDG values for $f_0(1500)$.
The largest decay channel should be $4\pi$ either through $\rho\rho$ directly or by its cross talk with $\sigma\sigma$.
Note that due to the binding energy of the molecule as well as the kinetic energy of $\rho$ inside the molecule, the $4\pi$ decay width through the decay of $\rho$ inside the $\rho\rho$ molecule can be smaller than the decay width of a single free $\rho$ meson.  Similar effect was pointed out by Refs.~\cite{Niskanen:2016ntu,Gal:2016bhp} in their studies of $d^*(2380)$ as a $\Delta\Delta$ molecule which gets a decay width smaller than the decay width of a single free $\Delta$ state. This kind of effect was also observed by the study of other hadronic molecules~\cite{Lin:2018nqd,Lin:2018kcc}.

In summary, the $\rho\rho$ scattering is revisited by including its coupled-channels of pseudoscalar mesons, {\sl i.e.}, $\pi\pi$ and $\bar KK$. It is found that the coupled-channel effect is important and shifts up the pole mass of the dynamically generated scalar state significantly. It makes the state to be more consistent with $f_0(1500)$ rather than $f_0(1370)$ as favored by the previous studies~\cite{Molina:2008jw,Gulmez:2016scm} without including these coupled channels.
This leads to a nicely consistent picture with a recent dispersive study~\cite{Ropertz:2018stk}
where a new parametrization for the scalar pion form factors is derived by fitting it to LHCb data on $\bar B ^0_s \to J/\psi \pi \pi$, and find an $f_0(1500)$
at mass $1465\pm 18 MeV$ coupling strongly to $\rho \rho$ (or $\sigma \sigma$).
The $\rho\rho$ scattering has been extended to the S-wave interactions for the whole vector-meson nonet by two groups~\cite{Geng:2008gx,Du:2018gyn}. Both propose $f_0(1710)$ to be $K^*\bar K^*$ dynamically generated state.  We expect similar significant coupled channel effects there. By including its coupled-channels of pseudoscalar mesons, the $K^*\bar K^*$ dynamically generated state could be $f_0(1790)$ suggested by the BES data~\cite{Ablikim:2004wn,Ablikim:2004qna,Ablikim:2004st} instead of $f_0(1710)$ .
The $f_0(1370)$, $f_0(1500)$ and $f_0(1710)$ have been studied before in quarkonia-glueball mixing picture in Refs.~\cite{Close:2005vf,Giacosa:2005zt,Chatzis:2011qz}, trying to
pin down partial contributions of glueball, nonstrange and strange quakonia in these scalar mesons. With the new configuration of meson-meson dynamically generated states, the structure of these scalars should be richer than previous assumptions and deserve further exploration by expanding the configuration space.

\section*{Acknowledgments}

We thank Li-Sheng Geng, Feng-Kun Guo, Ulf-G. Meissner and Eulogio Oset for helpful discussions. This project is supported by NSFC under Grant No.~11621131001 (CRC110 cofunded by DFG and NSFC) and Grant No.~11835015.


\end{document}